\documentclass{article}

\usepackage{microtype}
\usepackage{graphicx}
\usepackage{subfigure}
\usepackage{booktabs} 

\usepackage{hyperref}

\usepackage[accepted]{icml2023}

\usepackage{amsmath}
\usepackage{amssymb}
\usepackage{mathtools}
\usepackage{amsthm}

\usepackage[capitalize,noabbrev]{cleveref}

\theoremstyle{plain}

\theoremstyle{definition}

\theoremstyle{remark}

\usepackage[textsize=tiny]{todonotes}

\icmltitlerunning{Toward a Spectral Foundation Model}

\begin{document}

\twocolumn[
\icmltitle{Toward a Spectral Foundation Model: An Attention-Based Approach with Domain-Inspired Fine-Tuning and Wavelength Parameterization}




\begin{icmlauthorlist}
\icmlauthor{Tomasz R\'o\.za\'nski\textsuperscript{ *}}{wroclaw,rsaa}
\icmlauthor{Yuan-Sen Ting\textsuperscript{ *}}{rsaa,soco,osu}
\icmlauthor{Maja Jab\l{}o\'nska}{rsaa,warsaw}
\end{icmlauthorlist}

\icmlaffiliation{wroclaw}{Astronomical Institute, University of Wroc\l aw, Kopernika 11, 51-622, Wroc\l aw, Poland}
\icmlaffiliation{rsaa}{Research School of Astronomy \& Astrophysics, Australian National University, Cotter Rd., Weston, ACT 2611, Australia}
\icmlaffiliation{soco}{School of Computing, Australian National University, Acton, ACT 2601, Australia}
\icmlaffiliation{osu}{Department of Astronomy, The Ohio State University, Columbus, USA}
\icmlaffiliation{warsaw}{Astronomical Observatory of University of Warsaw, Warsaw, Poland}

\icmlcorrespondingauthor{Tomasz R\'o\.za\'nski}{tomasz.rozanski@uwr.edu.pl}
\icmlcorrespondingauthor{Yuan-Sen Ting}{yuan-sen.ting@anu.edu.au}

\icmlkeywords{Machine Learning, ICML}

\vskip 0.3in
]



\printAffiliationsAndNotice{\icmlEqualContribution} 

\begin{abstract}
Astrophysical explorations are underpinned by large-scale stellar spectroscopy surveys, necessitating a paradigm shift in spectral fitting techniques. Our study proposes three enhancements to transcend the limitations of the current spectral emulation models. We implement an attention-based emulator, adept at unveiling long-range information between wavelength pixels. We leverage a domain-specific fine-tuning strategy where the model is pre-trained on spectra with fixed stellar parameters and variable elemental abundances, followed by fine-tuning on the entire domain. Moreover, by treating wavelength as an autonomous model parameter, akin to neural radiance fields, the model can generate spectra on any wavelength grid. In the case with a training set of $\mathcal{O}(1000)$, our approach exceeds current leading methods by a factor of 5-10 across all metrics.
\end{abstract}

\section{Introduction}

Spectroscopy holds a pivotal role in astrophysics, granting us the ability to decipher the properties of celestial bodies, such as stars and galaxies. Recent advancements in stellar spectroscopy, driven by large-scale surveys such as APOGEE, LAMOST, Gaia-ESO, and GALAH \citep{Gilmore2012,Luo2015,Majewski2017,Buder2020}, have necessitated a radical reassessment of our analytic methods to accommodate the surge in high-quality spectra, from a few thousand \citep{Fuhrmann1998,Bensby2003} to millions. This burgeoning data volume has led to the development of extensive grids of stellar spectra.

However, modelling stellar spectra, rich with crucial information about effective temperature, surface gravity, atmospheric velocity fields, and individual element abundances, presents significant computational challenges. This is particularly true when dealing with 3D models and non-local thermodynamic equilibrium (non-LTE) effects \citep{Amarsi2020,Bergemann2021,Gerber2023}. Machine learning-based emulators have shown promise in offering an improvement over traditional polynomial interpolation techniques in terms of sample efficiency \citep{Fabbro2017,Leung2018,OBriain2021}.  

Spectral emulators like The Cannon \citep{the_cannon} and The Payne \citep{the_payne,Straumit2022,Xiang2022} have been developed and adopted for large-scale spectroscopic surveys. Yet, these methods carry their own shortcomings. Both The Cannon, based on Ridge regression, fall short in expressiveness, The Payne, which relies on a neural network-based multilayer perceptron, does not impose sufficient inductive bias and hence its performance plateaus with large training data. Even though these models show adeptness in emulating global spectra, they usually carry an error margin of about 1\%, which hampers the detection of weaker signals, including subtler features, such as hyperfine structures, stellar rotation, pulsations, ultra-high-resolution radial velocity measurements, or starspots, underscoring the urgency for innovative amortization strategies. Furthermore, most if not all spectral emulation techniques proposed have often trained on a fixed wavelength grid, thereby restricting their applicability to datasets with different sampling wavelengths.

To overcome these challenges, we propose a novel solution: an attention-based emulator \citep{Vaswani2017,Shaw2018,Wang2020} for stellar spectra. We employ a transformer model \citep{Parmar2018,Han2021}, initially pre-trained on spectra with fixed stellar parameters and variable elemental abundances, and then fine-tuned with other parameters. This method significantly enhances spectral emulation accuracy and sample efficiency, surpassing established state-of-the-art spectral emulation techniques. Furthermore, inspired by neural radiance fields \citep{Mildenhall2020,Sitzmann2020,Sajjadi2021}, we treat wavelength as a distinct model parameter. This approach allows us to generate spectra on any wavelength grid with a single training iteration, thus removing a significant limitation in the current spectral emulation landscape.
\begin{figure}[!ht]
\label{fig:transformer}
\vskip 0.2in
\begin{center}
\centerline{\includegraphics[width=0.75\columnwidth]{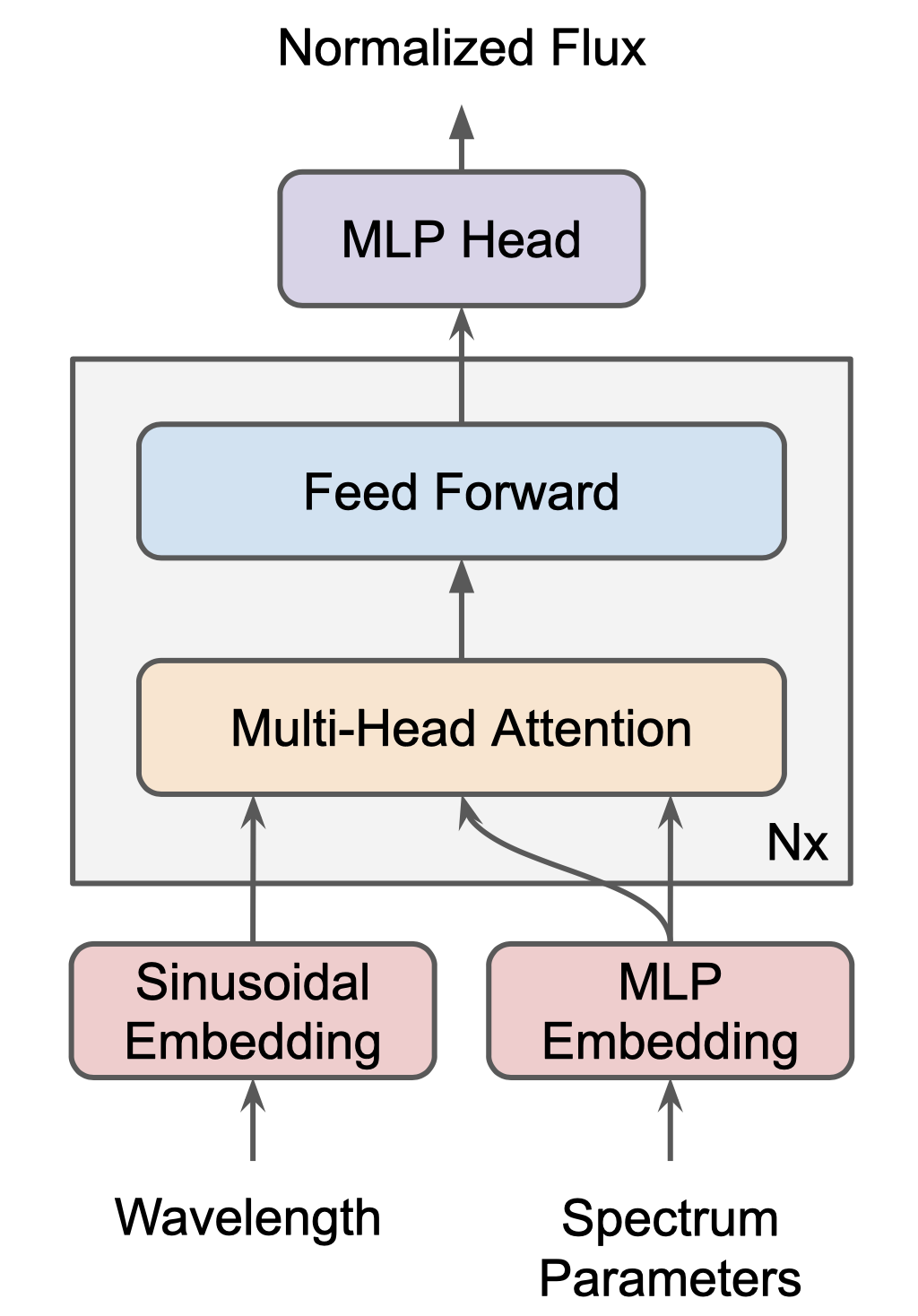}}
\caption{Architecture of the Proposed Transformer-Based Stellar Spectra Emulator: The model accepts two inputs: wavelength and a vector of spectrum parameters. Wavelength is encoded into a query token via sinusoidal encoding, while the parameters are transformed into tokens using an MLP Embedding. Transformer blocks capture long-range information, and the normalized flux is computed using an MLP Head.}
\label{fig1}
\end{center}
\vskip -0.2in
\end{figure}

\begin{figure*}[ht]
\vskip 0.2in
\begin{center}
\centerline{\includegraphics[width=0.95\linewidth]{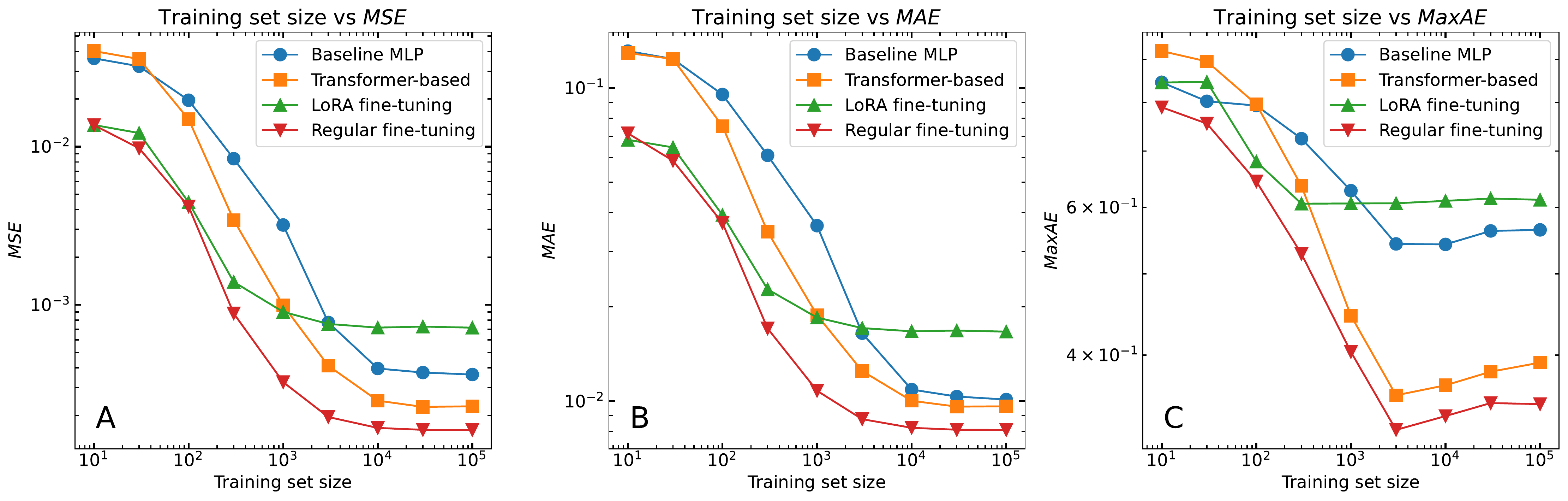}}
\caption{Assessment of our model's performance, illustrating the three essential metrics. Each panel showcases the performance of the baseline MLP, the Transformer-based model, and two variants of the fine-tuned Transformer-based models: LoRA fine-tuning and full fine-tuning. The Transformer-based model consistently outperforms the baseline MLP, with fine-tuning further enhancing the results, surpassing the state-of-the-art by a large margin in all metrics. The full fine-tuning approach demonstrates superior performance compared to the parameter-efficient fine-tuning method, LoRA, whose quality plateaus when the number of training examples $>1000$.}
\label{fig2}
\end{center}
\vskip -0.2in
\end{figure*}

\section{Synthetic Spectra}
In our experiments, we utilized two grids of synthetic spectra calculated using the plane-parallel model atmosphere codes updated by \citet{2008A&A...491..633L}. These updated codes are based on the standard Local Thermodynamic Equilibrium (LTE) codes ATLAS/SYNTHE \citep{1979ApJS...40....1K, 1993KurCD..18.....K,Kurucz2005,Kurucz2013}. Each grid consisted of 100,000 spectra calculated at a resolution of $R=100,000$ within the 4000 to 5000 Å wavelength range.

One key insight we will demonstrate is the transformers' ability to capture long-range information, particularly evident when stellar parameters are fixed. In such cases, atomic features associated with specific spectral lines become more prominent. By pre-training the model on synthetic spectra with fixed stellar parameters, allows the model to learn long-range correlations associated with individual atomic features. The second grid represented the comprehensive target domain, where we aimed to leverage pre-training for enhanced performance. We anticipate improved performance through fine-tuning the transformer model.

The first "pre-training" grid was established with a fixed effective temperature of $5000$ K, an effective gravity of $\log g=4.5$, a microturbulent velocity of $\xi=0$ km/s, and helium abundance ranging from zero to twice the solar value. Other element abundances were uniformly sampled between -2 and 1 dex. For the second ``entire" grid, we sampled the effective temperature from 4000 K to 6000 K and the effective gravity between 4.0 and 5.0, while keeping the other parameters identical to the first grid. Table \ref{tab:grids} provides a summary of the grids. The elemental abundances of all elements up to atomic number equal 98 were considered.

We emphasize that, while we tested our models on the Local Thermodynamic Equilibrium (LTE) model, the primary objective was to evaluate how the model performance scales with the volume of training data using different strategies. Our proposed model demonstrates robust emulation capabilities even with limited training data, facilitating the transition to amortizing non-LTE models.
\begin{table}
\label{tab:grids}
\centering
\caption{Grids of synthetic spectra adopted in this study.}
\begin{tabular}{lcc}
\hline
Stellar Labels & First Grid & Second Grid \\
\hline
Effective Temperature & 5000 K & $[4000,6000]$ K \\
Surface Gravity & 4.5 & $[4.0,5.0]$ \\
\# Training Spectra & \multicolumn{2}{c}{Up to 
 100,000} \\
\# Frequencies & \multicolumn{2}{c}{22,315} \\
Microturbulence & \multicolumn{2}{c}{0 km/s} \\
Helium Abundance & \multicolumn{2}{c}{$[0,0.1568]$} \\
Other Abundances ${\rm [X/H]}$ & \multicolumn{2}{c}{$[-2,1]$} \\
\hline
\end{tabular}
\end{table}

\section{Method}

The objective of our research was to develop a stellar spectra emulator that approximates the complex numerical modeling function $f(\lambda,\phi)$, where $\lambda$ represents the wavelength and $\phi$ denotes a vector of variable parameters. Our goal was to improve this function by minimizing the mean squared error (MSE) loss over a training dataset $\{(\lambda,\phi)_i,y_i\}_N$. As a baseline, we used the commonly employed Multilayer Perceptron (MLP) approach \citep{Straumit2022,Xiang2022} which predicts fluxes at fixed wavelengths. Our model stands apart through three core improvements.

{\bf 1. Transformer-based Model:} Traditional MLPs have been the go-to choice in the field, but they have their limitations. Stellar spectra, with its intrinsic complexity, have proved challenging for astro-ML practitioners, particularly due to the lack of discernible inductive biases. While Convolutional Neural Networks (CNNs) assume translational invariance and distortion stability, these biases don't apply to stellar spectra. Positional information (wavelength, in our case) is critical and spectral feature distortions can reveal important details about the gravitational broadening of stars. It is crucial that these factors aren't disregarded. Our thorough ablation studies have confirmed that CNNs do not perform better than our baseline MLP for these tasks, as the strong but imprecise inductive biases hinder spectra emulation. Instead, we've used a Transformer-based model which, unlike MLPs and CNNs, excels at capturing long-range information, a valuable trait for spectral emulation.

The transformer blocks play a crucial role in capturing the long-range information embedded within the spectra. We employ multi-head attention mechanisms within these transformer blocks to allow the model to attend to different parts of the multi-token embedding of the spectrum parameters simultaneously. This makes it straightforward for the model to establish complex correlations between distant wavelengths by attending to relevant tokens. The ability to learn a rich correlation structure is built into the model architecture. The residual connections and Layer Normalization are positioned following the recommendations by \citet{2023arXiv230414802X}. Our transformer-based model consists of 16 transformer blocks, with the output of the final block passed to an MLP Head. The MLP Head comprises two layers: an initial layer with 256 neurons and a final layer with a single neuron responsible for predicting the normalized flux. This structure facilitates a more comprehensive understanding of stellar spectra, leveraging the attention mechanisms of the transformers. The detailed architecture of the proposed transformer-based stellar spectra emulator is illustrated in Figure~\ref{fig1}.

{\bf 2. Two-Step Pretraining Strategy:} Recognizing that long-range information is particularly valuable at fixed stellar parameters, we adopted a two-step pretraining strategy. Post-pretraining, we've used two distinct fine-tuning strategies: full fine-tuning, where all model parameters are adjusted during target domain training, and a parameter-efficient fine-tuning method named LoRA \citep{Hu2021}, which is designed for fine-tuning large-scale language models when full fine-tuning is computationally prohibitive.

{\bf 3. Flexible Wavelength Output:} Finally, our model allows for the inclusion of wavelength as an input. The wavelength input is encoded into a singular token using sinusoidal encoding, which serves as a unique query token. Meanwhile, the vector of parameters is transformed into a series of tokens using an MLP Embedding, acting as keys and values tokens in subsequent transformer blocks. This means, without extra training, a single trained model can interpret spectra that are sampled at different wavelength grids, thus addressing a significant bottleneck in spectral emulation. 

{\bf Metrics:} To obtain a comprehensive evaluation of our model's performance, we employed an array of key metrics including Mean Squared Error (MSE), Mean Absolute Error (MAE) and Maximum Absolute Error (MaxAE). Although not shown, we also checked the Mean Absolute Relative Error (MARE), the 95th percentile of absolute error (${QAE}_{0.95}$), and the mean of the top 5\% absolute errors (${QMAE}_{0.95}$). Our decision to employ these diverse metrics stems from our understanding that using a singular measure like MSE could potentially provide a misleading view of our emulator's performance. An inaccurate emulation of key spectral features can considerably influence the quality of our inferences, even if the emulator correctly represents most of the less informative spectral continuum. Our chosen metrics offer a more balanced evaluation of our model's performance.

{\bf Training:} We've employed the AdamW optimizer with a cosine learning rate schedule. Training was conducted on Grid 2 for 100,000 steps for both the Baseline MLP and the Transformer-based models, pre-training ran for 100,000 steps on Grid 1, and fine-tuning was carried out for 50,000 steps for both full fine-tuning and LoRA. The pre-training was conducted on 4 A100 GPUs and took approximately 16 hours, with a batch size of 64. All other training sessions were run on a single GPU with a batch size of 16.
\begin{figure}[!ht]
\label{fig:Fig3}
\begin{center}
\centerline{\includegraphics[width=0.9\columnwidth]{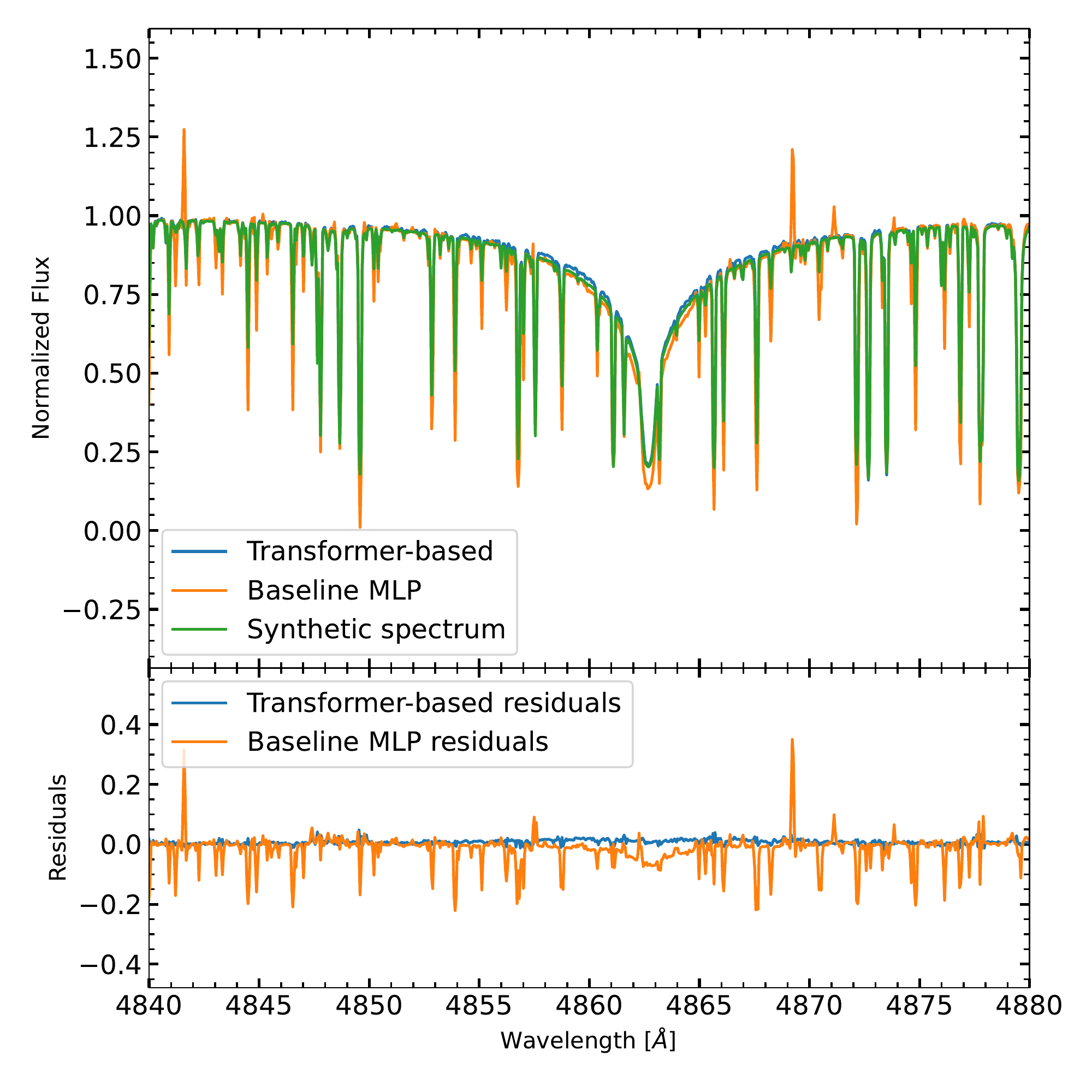}}
\vskip -0.1in
\caption{This visual illustrates how the fine-tuned Transformer, despite sparse, high-dimensional data (ndim $= 100$ with all elemental abundances), replicates a complex synthetic spectrum with remarkable precision, unlike the struggling baseline MLP model, with both models trained on a dataset of 1000 spectra.}
\label{fig3}
\end{center}
\vskip -0.3in
\end{figure}

\section{Results}

The results of our experiments, as shown in Fig~\ref{fig2}, demonstrate a significant improvement in performance when comparing the baseline MLP (blue) to the transformer-based model (orange) across all metrics. The transformer-based model consistently achieves two-to-three times better performance, even for metrics such as MaxAE, which evaluate the emulation of strong features where important spectral information resides. Importantly, unlike MLP models that train on a fixed wavelength grid, our transformer-based model provides the flexibility to emulate spectra on any wavelength grid within the training data's wavelength range, with no degradation of the performance.

Notably, the difference between the MLP and transformer models becomes more pronounced when the training set comprises between 100 and 3000 spectra. In this range, the transformer-based model often achieves comparable results with approximately three times fewer training examples

Our domain-inspired fine-tuning strategy yields another 2-3 times improvement compared to the basic transformer models. This translates to a 5-10 fold enhancement across all metrics, with a training size of 1000, when compared to current leading methods in the field. (see Fig~\ref{fig3}) While we have employed a relatively modest transformer model in this study, it is worth highlighting that we have not yet reached a saturation point in terms of the number of units/layers for the transformers. This suggests that with appropriate inductive bias, such as assuming that atomic features convey long-range information in the spectra, the models can continue to improve with increasing computational power. When comparing full fine-tuning to MLP, we typically require only 1/10 of the spectra to achieve similar results.

In addition to full fine-tuning, we also explore a low-cost fine-tuning strategy called LoRA, which is commonly used for large-language models \citep{Hu2021}. As shown in Fig~\ref{fig2}, LoRA demonstrates the potential to perform on par with full fine-tuning models with fewer than a few hundred training examples. However, its performance becomes constrained when more spectra are available, making the full fine-tuning more favourable in this context. Nonetheless, this intriguing result suggests that even with the release of larger foundational spectral models, research groups with limited GPU resources and access to a few hundred to thousand high-quality 3D non-LTE models can still benefit from the LoRA-adapter type of fine-tuning while leveraging our spectral foundation model.

\section{Future Direction and Broader Impact}

Our investigation has demonstrated the superiority of Transformer-based models over the state-of-the-art model, particularly when employing domain-inspired fine-tuning strategies. However, we recognize that there is still room for further exploration and improvement. We aim to delve deeper into the limitations of adapter methods \citep{He2021,Karimi2021,Hu2021} when confronted with a substantial number of training examples. Moreover, our vision extends to expanding our study by incorporating more complex downstream supervised tasks, such as simultaneous prediction of spectral fluxes and stellar atmospheric structures. This integration holds immense potential for guiding the training process more effectively and unlocking even better results.

Looking beyond our specific findings, there is a growing demand for a generalizable and robust algorithm that can be seamlessly transferred and adapted across different spectroscopic surveys. Presently, each individual survey invests significant effort in constructing its own pipeline and developing unique spectral emulators. We propose that the foundation model could be initially trained on a dataset encompassing millions of simpler (e.g., 1D LTE) high-resolution spectra, and then fine-tuned with thousands of spectra derived from more sophisticated physics models or those specifically tailored for a particular survey. Our work represents an important initial step towards the development of a genuinely robust, adjustable, and adaptable spectral foundation model. Such a model would possess the remarkable capability of generating spectra on any wavelength grid, providing a unifying platform for the field. Drawing parallels to the impact of large language models, we firmly believe that these evolving generic transformer-based methods will revolutionize the spectroscopic domain. They will establish a common ground for all studies, empowering researchers to fine-tune their preferred spectral models without starting from scratch. This will not only enhance efficiency and collaboration within the community but also enable exciting new possibilities for scientific exploration and discovery.

\bibliography{paper}

\begin{thebibliography}{33}
\providecommand{\natexlab}[1]{#1}
\providecommand{\url}[1]{\texttt{#1}}
\expandafter\ifx\csname urlstyle\endcsname\relax
  \providecommand{\doi}[1]{doi: #1}\else
  \providecommand{\doi}{doi: \begingroup \urlstyle{rm}\Url}\fi

\bibitem[{Amarsi} et~al.(2020){Amarsi}, {Lind}, {Osorio}, {Nordlander},
  {Bergemann}, {Reggiani}, {Wang}, {Buder}, {Asplund}, {Barklem}, {Wehrhahn},
  {Sk{\'u}lad{\'o}ttir}, {Kobayashi}, {Karakas}, {Gao}, {Bland-Hawthorn}, {de
  Silva}, {Kos}, {Lewis}, {Martell}, {Sharma}, {Simpson}, {Zucker},
  {{\v{C}}otar}, {Horner}, and {GALAH Collaboration}]{Amarsi2020}
{Amarsi}, A.~M., {Lind}, K., {Osorio}, Y., {Nordlander}, T., {Bergemann}, M.,
  {Reggiani}, H., {Wang}, E.~X., {Buder}, S., {Asplund}, M., {Barklem}, P.~S.,
  {Wehrhahn}, A., {Sk{\'u}lad{\'o}ttir}, {\'A}., {Kobayashi}, C., {Karakas},
  A.~I., {Gao}, X.~D., {Bland-Hawthorn}, J., {de Silva}, G.~M., {Kos}, J.,
  {Lewis}, G.~F., {Martell}, S.~L., {Sharma}, S., {Simpson}, J.~D., {Zucker},
  D.~B., {{\v{C}}otar}, K., {Horner}, J., and {GALAH Collaboration}.
\newblock {The GALAH Survey: non-LTE departure coefficients for large
  spectroscopic surveys}.
\newblock \emph{\aap}, 642:\penalty0 A62, October 2020.
\newblock \doi{10.1051/0004-6361/202038650}.

\bibitem[{Bensby} et~al.(2003){Bensby}, {Feltzing}, and
  {Lundstr{\"o}m}]{Bensby2003}
{Bensby}, T., {Feltzing}, S., and {Lundstr{\"o}m}, I.
\newblock {Elemental abundance trends in the Galactic thin and thick disks as
  traced by nearby F and G dwarf stars}.
\newblock \emph{\aap}, 410:\penalty0 527--551, November 2003.
\newblock \doi{10.1051/0004-6361:20031213}.

\bibitem[Bergemann et~al.(2021)Bergemann, Hoppe, Semenova, Carlsson, Yakovleva,
  Voronov, Bautista, Nemer, Belyaev, Leenaarts, Mashonkina, Reiners, and
  Ellwarth]{Bergemann2021}
Bergemann, M., Hoppe, R., Semenova, E., Carlsson, M., Yakovleva, S.~A.,
  Voronov, Y.~V., Bautista, M., Nemer, A., Belyaev, A.~K., Leenaarts, J.,
  Mashonkina, L., Reiners, A., and Ellwarth, M.
\newblock Solar oxygen abundance.
\newblock \emph{Monthly Notices of the Royal Astronomical Society},
  508\penalty0 (2):\penalty0 2236--2253, jul 2021.
\newblock \doi{10.1093/mnras/stab2160}.
\newblock URL \url{https://doi.org/10.1093%2Fmnras%2Fstab2160}.

\bibitem[{Buder} et~al.(2020){Buder}, {Sharma}, {Kos}, {Amarsi}, {Nordlander},
  {Lind}, {Martell}, {Asplund}, {Bland-Hawthorn}, {Casey}, {De Silva},
  {D'Orazi}, {Freeman}, {Hayden}, {Lewis}, {Lin}, {Schlesinger}, {Simpson},
  {Stello}, {Zucker}, {Zwitter}, {Beeson}, {Buck}, {Casagrande}, {Clark},
  {Cotar}, {Da Costa}, {de Grijs}, {Feuillet}, {Horner}, {Khanna}, {Kafle},
  {Liu}, {Montet}, {Nandakumar}, {Nataf}, {Ness}, {Spina}, {Traven},
  {Tepper-Garcia}, {Ting}, {Vogrincic}, {Wittenmyer}, {Zerjal}, and {the GALAH
  collaboration}]{Buder2020}
{Buder}, S., {Sharma}, S., {Kos}, J., {Amarsi}, A.~M., {Nordlander}, T.,
  {Lind}, K., {Martell}, S.~L., {Asplund}, M., {Bland-Hawthorn}, J., {Casey},
  A.~R., {De Silva}, G.~M., {D'Orazi}, V., {Freeman}, K.~C., {Hayden}, M.~R.,
  {Lewis}, G.~F., {Lin}, J., {Schlesinger}, K.~J., {Simpson}, J.~D., {Stello},
  D., {Zucker}, D.~B., {Zwitter}, T., {Beeson}, K.~L., {Buck}, T.,
  {Casagrande}, L., {Clark}, J.~T., {Cotar}, K., {Da Costa}, G.~S., {de Grijs},
  R., {Feuillet}, D., {Horner}, J., {Khanna}, S., {Kafle}, P.~R., {Liu}, F.,
  {Montet}, B.~T., {Nandakumar}, G., {Nataf}, D.~M., {Ness}, M.~K., {Spina},
  L., {Traven}, G., {Tepper-Garcia}, T., {Ting}, Y.-S., {Vogrincic}, R.,
  {Wittenmyer}, R.~A., {Zerjal}, M., and {the GALAH collaboration}.
\newblock {The GALAH+ Survey: Third Data Release}.
\newblock \emph{arXiv e-prints}, art. arXiv:2011.02505, November 2020.

\bibitem[Fabbro et~al.(2017)Fabbro, Venn, O{\textquotesingle}Briain, Bialek,
  Kielty, Jahandar, and Monty]{Fabbro2017}
Fabbro, S., Venn, K.~A., O{\textquotesingle}Briain, T., Bialek, S., Kielty,
  C.~L., Jahandar, F., and Monty, S.
\newblock An application of deep learning in the analysis of stellar spectra.
\newblock \emph{Monthly Notices of the Royal Astronomical Society},
  475\penalty0 (3):\penalty0 2978--2993, dec 2017.
\newblock \doi{10.1093/mnras/stx3298}.
\newblock URL \url{https://doi.org/10.1093%2Fmnras%2Fstx3298}.

\bibitem[{Fuhrmann}(1998)]{Fuhrmann1998}
{Fuhrmann}, K.
\newblock {Nearby stars of the Galactic disk and halo}.
\newblock \emph{\aap}, 338:\penalty0 161--183, October 1998.

\bibitem[Gerber et~al.(2023)Gerber, Magg, Plez, Bergemann, Heiter, Olander, and
  Hoppe]{Gerber2023}
Gerber, J.~M., Magg, E., Plez, B., Bergemann, M., Heiter, U., Olander, T., and
  Hoppe, R.
\newblock Non-{LTE} radiative transfer with turbospectrum.
\newblock \emph{Astronomy and Astrophysics}, 669:\penalty0 A43, jan 2023.
\newblock \doi{10.1051/0004-6361/202243673}.
\newblock URL \url{https://doi.org/10.1051%2F0004-6361%2F202243673}.

\bibitem[{Gilmore} et~al.(2012){Gilmore}, {Randich}, {Asplund}, {Binney},
  {Bonifacio}, {Drew}, {Feltzing}, {Ferguson}, {Jeffries}, {Micela},
  {Negueruela}, {Prusti}, {Rix}, {Vallenari}, {Alfaro}, {Allende-Prieto},
  {Babusiaux}, {Bensby}, {Blomme}, {Bragaglia}, {Flaccomio}, {Fran{\c{c}}ois},
  {Irwin}, {Koposov}, {Korn}, {Lanzafame}, {Pancino}, {Paunzen},
  {Recio-Blanco}, {Sacco}, {Smiljanic}, {Van Eck}, {Walton}, {Aden}, {Aerts},
  {Affer}, {Alcala}, {Altavilla}, {Alves}, {Antoja}, {Arenou}, {Argiroffi},
  {Asensio Ramos}, {Bailer-Jones}, {Balaguer-Nunez}, {Bayo}, {Barbuy},
  {Barisevicius}, {Barrado y Navascues}, {Battistini}, {Bellas Velidis},
  {Bellazzini}, {Belokurov}, {Bergemann}, {Bertelli}, {Biazzo}, {Bienayme},
  {Bland-Hawthorn}, {Boeche}, {Bonito}, {Boudreault}, {Bouvier}, {Brandao},
  {Brown}, {de Bruijne}, {Burleigh}, {Caballero}, {Caffau}, {Calura},
  {Capuzzo-Dolcetta}, {Caramazza}, {Carraro}, {Casagrande}, {Casewell},
  {Chapman}, {Chiappini}, {Chorniy}, {Christlieb}, {Cignoni}, {Cocozza},
  {Colless}, {Collet}, {Collins}, {Correnti}, {Covino}, {Crnojevic}, {Cropper},
  {Cunha}, {Damiani}, {David}, {Delgado}, {Duffau}, {Edvardsson}, {Eldridge},
  {Enke}, {Eriksson}, {Evans}, {Eyer}, {Famaey}, {Fellhauer}, {Ferreras},
  {Figueras}, {Fiorentino}, {Flynn}, {Folha}, {Franciosini}, {Frasca},
  {Freeman}, {Fremat}, {Friel}, {Gaensicke}, {Gameiro}, {Garzon}, {Geier},
  {Geisler}, {Gerhard}, {Gibson}, {Gomboc}, {Gomez}, {Gonzalez-Fernandez},
  {Gonzalez Hernandez}, {Gosset}, {Grebel}, {Greimel}, {Groenewegen},
  {Grundahl}, {Guarcello}, {Gustafsson}, {Hadrava}, {Hatzidimitriou}, {Hambly},
  {Hammersley}, {Hansen}, {Haywood}, {Heber}, {Heiter}, {Held}, {Helmi},
  {Hensler}, {Herrero}, {Hill}, {Hodgkin}, {Huelamo}, {Huxor}, {Ibata},
  {Jackson}, {de Jong}, {Jonker}, {Jordan}, {Jordi}, {Jorissen}, {Katz},
  {Kawata}, {Keller}, {Kharchenko}, {Klement}, {Klutsch}, {Knude}, {Koch},
  {Kochukhov}, {Kontizas}, {Koubsky}, {Lallement}, {de Laverny}, {van Leeuwen},
  {Lemasle}, {Lewis}, {Lind}, {Lindstrom}, {Lobel}, {Lopez Santiago}, {Lucas},
  {Ludwig}, {Lueftinger}, {Magrini}, {Maiz Apellaniz}, {Maldonado}, {Marconi},
  {Marino}, {Martayan}, {Martinez-Valpuesta}, {Matijevic}, {McMahon},
  {Messina}, {Meyer}, {Miglio}, {Mikolaitis}, {Minchev}, {Minniti}, {Moitinho},
  {Momany}, {Monaco}, {Montalto}, {Monteiro}, {Monier}, {Montes}, {Mora},
  {Moraux}, {Morel}, {Mowlavi}, {Mucciarelli}, {Munari}, {Napiwotzki},
  {Nardetto}, {Naylor}, {Naze}, {Nelemans}, {Okamoto}, {Ortolani}, {Pace},
  {Palla}, {Palous}, {Parker}, {Penarrubia}, {Pillitteri}, {Piotto}, {Posbic},
  {Prisinzano}, {Puzeras}, {Quirrenbach}, {Ragaini}, {Read}, {Read}, {Reyle},
  {De Ridder}, {Robichon}, {Robin}, {Roeser}, {Romano}, {Royer}, {Ruchti},
  {Ruzicka}, {Ryan}, {Ryde}, {Santos}, {Sanz Forcada}, {Sarro Baro},
  {Sbordone}, {Schilbach}, {Schmeja}, {Schnurr}, {Schoenrich}, {Scholz},
  {Seabroke}, {Sharma}, {De Silva}, {Smith}, {Solano}, {Sordo}, {Soubiran},
  {Sousa}, {Spagna}, {Steffen}, {Steinmetz}, {Stelzer}, {Stempels},
  {Tabernero}, {Tautvaisiene}, {Thevenin}, {Torra}, {Tosi}, {Tolstoy}, {Turon},
  {Walker}, {Wambsganss}, {Worley}, {Venn}, {Vink}, {Wyse}, {Zaggia},
  {Zeilinger}, {Zoccali}, {Zorec}, {Zucker}, {Zwitter}, and {Gaia-ESO Survey
  Team}]{Gilmore2012}
{Gilmore}, G., {Randich}, S., {Asplund}, M., {Binney}, J., {Bonifacio}, P.,
  {Drew}, J., {Feltzing}, S., {Ferguson}, A., {Jeffries}, R., {Micela}, G.,
  {Negueruela}, I., {Prusti}, T., {Rix}, H.~W., {Vallenari}, A., {Alfaro}, E.,
  {Allende-Prieto}, C., {Babusiaux}, C., {Bensby}, T., {Blomme}, R.,
  {Bragaglia}, A., {Flaccomio}, E., {Fran{\c{c}}ois}, P., {Irwin}, M.,
  {Koposov}, S., {Korn}, A., {Lanzafame}, A., {Pancino}, E., {Paunzen}, E.,
  {Recio-Blanco}, A., {Sacco}, G., {Smiljanic}, R., {Van Eck}, S., {Walton},
  N., {Aden}, D., {Aerts}, C., {Affer}, L., {Alcala}, J.~M., {Altavilla}, G.,
  {Alves}, J., {Antoja}, T., {Arenou}, F., {Argiroffi}, C., {Asensio Ramos},
  A., {Bailer-Jones}, C., {Balaguer-Nunez}, L., {Bayo}, A., {Barbuy}, B.,
  {Barisevicius}, G., {Barrado y Navascues}, D., {Battistini}, C., {Bellas
  Velidis}, I., {Bellazzini}, M., {Belokurov}, V., {Bergemann}, M., {Bertelli},
  G., {Biazzo}, K., {Bienayme}, O., {Bland-Hawthorn}, J., {Boeche}, C.,
  {Bonito}, S., {Boudreault}, S., {Bouvier}, J., {Brandao}, I., {Brown}, A.,
  {de Bruijne}, J., {Burleigh}, M., {Caballero}, J., {Caffau}, E., {Calura},
  F., {Capuzzo-Dolcetta}, R., {Caramazza}, M., {Carraro}, G., {Casagrande}, L.,
  {Casewell}, S., {Chapman}, S., {Chiappini}, C., {Chorniy}, Y., {Christlieb},
  N., {Cignoni}, M., {Cocozza}, G., {Colless}, M., {Collet}, R., {Collins}, M.,
  {Correnti}, M., {Covino}, E., {Crnojevic}, D., {Cropper}, M., {Cunha}, M.,
  {Damiani}, F., {David}, M., {Delgado}, A., {Duffau}, S., {Edvardsson}, B.,
  {Eldridge}, J., {Enke}, H., {Eriksson}, K., {Evans}, N.~W., {Eyer}, L.,
  {Famaey}, B., {Fellhauer}, M., {Ferreras}, I., {Figueras}, F., {Fiorentino},
  G., {Flynn}, C., {Folha}, D., {Franciosini}, E., {Frasca}, A., {Freeman}, K.,
  {Fremat}, Y., {Friel}, E., {Gaensicke}, B., {Gameiro}, J., {Garzon}, F.,
  {Geier}, S., {Geisler}, D., {Gerhard}, O., {Gibson}, B., {Gomboc}, A.,
  {Gomez}, A., {Gonzalez-Fernandez}, C., {Gonzalez Hernandez}, J., {Gosset},
  E., {Grebel}, E., {Greimel}, R., {Groenewegen}, M., {Grundahl}, F.,
  {Guarcello}, M., {Gustafsson}, B., {Hadrava}, P., {Hatzidimitriou}, D.,
  {Hambly}, N., {Hammersley}, P., {Hansen}, C., {Haywood}, M., {Heber}, U.,
  {Heiter}, U., {Held}, E., {Helmi}, A., {Hensler}, G., {Herrero}, A., {Hill},
  V., {Hodgkin}, S., {Huelamo}, N., {Huxor}, A., {Ibata}, R., {Jackson}, R.,
  {de Jong}, R., {Jonker}, P., {Jordan}, S., {Jordi}, C., {Jorissen}, A.,
  {Katz}, D., {Kawata}, D., {Keller}, S., {Kharchenko}, N., {Klement}, R.,
  {Klutsch}, A., {Knude}, J., {Koch}, A., {Kochukhov}, O., {Kontizas}, M.,
  {Koubsky}, P., {Lallement}, R., {de Laverny}, P., {van Leeuwen}, F.,
  {Lemasle}, B., {Lewis}, G., {Lind}, K., {Lindstrom}, H.~P.~E., {Lobel}, A.,
  {Lopez Santiago}, J., {Lucas}, P., {Ludwig}, H., {Lueftinger}, T., {Magrini},
  L., {Maiz Apellaniz}, J., {Maldonado}, J., {Marconi}, G., {Marino}, A.,
  {Martayan}, C., {Martinez-Valpuesta}, I., {Matijevic}, G., {McMahon}, R.,
  {Messina}, S., {Meyer}, M., {Miglio}, A., {Mikolaitis}, S., {Minchev}, I.,
  {Minniti}, D., {Moitinho}, A., {Momany}, Y., {Monaco}, L., {Montalto}, M.,
  {Monteiro}, M.~J., {Monier}, R., {Montes}, D., {Mora}, A., {Moraux}, E.,
  {Morel}, T., {Mowlavi}, N., {Mucciarelli}, A., {Munari}, U., {Napiwotzki},
  R., {Nardetto}, N., {Naylor}, T., {Naze}, Y., {Nelemans}, G., {Okamoto}, S.,
  {Ortolani}, S., {Pace}, G., {Palla}, F., {Palous}, J., {Parker}, R.,
  {Penarrubia}, J., {Pillitteri}, I., {Piotto}, G., {Posbic}, H., {Prisinzano},
  L., {Puzeras}, E., {Quirrenbach}, A., {Ragaini}, S., {Read}, J., {Read}, M.,
  {Reyle}, C., {De Ridder}, J., {Robichon}, N., {Robin}, A., {Roeser}, S.,
  {Romano}, D., {Royer}, F., {Ruchti}, G., {Ruzicka}, A., {Ryan}, S., {Ryde},
  N., {Santos}, N., {Sanz Forcada}, J., {Sarro Baro}, L.~M., {Sbordone}, L.,
  {Schilbach}, E., {Schmeja}, S., {Schnurr}, O., {Schoenrich}, R., {Scholz},
  R.~D., {Seabroke}, G., {Sharma}, S., {De Silva}, G., {Smith}, M., {Solano},
  E., {Sordo}, R., {Soubiran}, C., {Sousa}, S., {Spagna}, A., {Steffen}, M.,
  {Steinmetz}, M., {Stelzer}, B., {Stempels}, E., {Tabernero}, H.,
  {Tautvaisiene}, G., {Thevenin}, F., {Torra}, J., {Tosi}, M., {Tolstoy}, E.,
  {Turon}, C., {Walker}, M., {Wambsganss}, J., {Worley}, C., {Venn}, K.,
  {Vink}, J., {Wyse}, R., {Zaggia}, S., {Zeilinger}, W., {Zoccali}, M.,
  {Zorec}, J., {Zucker}, D., {Zwitter}, T., and {Gaia-ESO Survey Team}.
\newblock {The Gaia-ESO Public Spectroscopic Survey}.
\newblock \emph{The Messenger}, 147:\penalty0 25--31, March 2012.

\bibitem[Han et~al.(2021)Han, Xiao, Wu, Guo, Xu, and Wang]{Han2021}
Han, K., Xiao, A., Wu, E., Guo, J., Xu, C., and Wang, Y.
\newblock Transformer in transformer.
\newblock \emph{Advances in Neural Information Processing Systems},
  34:\penalty0 15908--15919, 2021.

\bibitem[He et~al.(2021)He, Liu, Ye, Tan, Ding, Cheng, Low, Bing, and
  Si]{He2021}
He, R., Liu, L., Ye, H., Tan, Q., Ding, B., Cheng, L., Low, J.-W., Bing, L.,
  and Si, L.
\newblock On the effectiveness of adapter-based tuning for pretrained language
  model adaptation.
\newblock \emph{arXiv preprint arXiv:2106.03164}, 2021.

\bibitem[{Hu} et~al.(2021){Hu}, {Shen}, {Wallis}, {Allen-Zhu}, {Li}, {Wang},
  {Wang}, and {Chen}]{Hu2021}
{Hu}, E.~J., {Shen}, Y., {Wallis}, P., {Allen-Zhu}, Z., {Li}, Y., {Wang}, S.,
  {Wang}, L., and {Chen}, W.
\newblock {LoRA: Low-Rank Adaptation of Large Language Models}.
\newblock \emph{arXiv e-prints}, art. arXiv:2106.09685, June 2021.
\newblock \doi{10.48550/arXiv.2106.09685}.

\bibitem[Karimi~Mahabadi et~al.(2021)Karimi~Mahabadi, Henderson, and
  Ruder]{Karimi2021}
Karimi~Mahabadi, R., Henderson, J., and Ruder, S.
\newblock Compacter: Efficient low-rank hypercomplex adapter layers.
\newblock \emph{Advances in Neural Information Processing Systems},
  34:\penalty0 1022--1035, 2021.

\bibitem[{Kurucz}(1993)]{1993KurCD..18.....K}
{Kurucz}, R.
\newblock {SYNTHE Spectrum Synthesis Programs and Line Data.}
\newblock \emph{SYNTHE Spectrum Synthesis Programs and Line Data. Kurucz CD-ROM
  No. 18. Cambridge}, 18, January 1993.

\bibitem[{Kurucz}(1979)]{1979ApJS...40....1K}
{Kurucz}, R.~L.
\newblock {Model atmospheres for G, F, A, B, and O stars.}
\newblock \emph{The Astrophysical Journal Supplement Series}, 40:\penalty0
  1--340, May 1979.
\newblock \doi{10.1086/190589}.

\bibitem[{Kurucz}(2005)]{Kurucz2005}
{Kurucz}, R.~L.
\newblock {ATLAS12, SYNTHE, ATLAS9, WIDTH9, et cetera}.
\newblock \emph{Memorie della Societa Astronomica Italiana Supplementi},
  8:\penalty0 14, 2005.

\bibitem[{Kurucz}(2013)]{Kurucz2013}
{Kurucz}, R.~L.
\newblock {ATLAS12: Opacity sampling model atmosphere program}.
\newblock Astrophysics Source Code Library, March 2013.

\bibitem[{Lester} \& {Neilson}(2008){Lester} and
  {Neilson}]{2008A&A...491..633L}
{Lester}, J.~B. and {Neilson}, H.~R.
\newblock {satlas: spherical versions of the atlas stellar atmosphere program}.
\newblock \emph{Astronomy and Astrophysics}, 491\penalty0 (2):\penalty0
  633--641, November 2008.
\newblock \doi{10.1051/0004-6361:200810578}.

\bibitem[Leung \& Bovy(2018)Leung and Bovy]{Leung2018}
Leung, H.~W. and Bovy, J.
\newblock Deep learning of multi-element abundances from high-resolution
  spectroscopic data.
\newblock \emph{Monthly Notices of the Royal Astronomical Society}, nov 2018.
\newblock \doi{10.1093/mnras/sty3217}.
\newblock URL \url{https://doi.org/10.1093%2Fmnras%2Fsty3217}.

\bibitem[{Luo} et~al.(2015){Luo}, {Zhao}, {Zhao}, {Deng}, {Liu}, {Jing},
  {Wang}, {Zhang}, {Shi}, {Cui}, {Chu}, {Li}, {Bai}, {Wu}, {Cai}, {Cao}, {Cao},
  {Carlin}, {Chen}, {Chen}, {Chen}, {Chen}, {Chen}, {Chen}, {Chen},
  {Christlieb}, {Chu}, {Cui}, {Dong}, {Du}, {Fan}, {Feng}, {Fu}, {Gao}, {Gong},
  {Gu}, {Guo}, {Han}, {He}, {Hou}, {Hou}, {Hou}, {Hu}, {Hu}, {Hu}, {Huo},
  {Jia}, {Jiang}, {Jiang}, {Jiang}, {Jin}, {Kong}, {Kong}, {Lei}, {Li}, {Li},
  {Li}, {Li}, {Li}, {Li}, {Li}, {Li}, {Li}, {Li}, {Li}, {Li}, {Liang}, {Lin},
  {Liu}, {Liu}, {Liu}, {Liu}, {Lu}, {Luo}, {Mao}, {Newberg}, {Ni}, {Qi}, {Qi},
  {Shen}, {Shi}, {Song}, {Song}, {Su}, {Su}, {Tang}, {Tao}, {Tian}, {Wang},
  {Wang}, {Wang}, {Wang}, {Wang}, {Wang}, {Wang}, {Wang}, {Wang}, {Wang},
  {Wang}, {Wang}, {Wang}, {Wang}, {Wang}, {Wang}, {Wang}, {Wang}, {Wang},
  {Wang}, {Wei}, {Wei}, {Wu}, {Wu}, {Wu}, {Wu}, {Xing}, {Xu}, {Xu}, {Xu},
  {Yan}, {Yang}, {Yang}, {Yang}, {Yang}, {Yao}, {Yu}, {Yuan}, {Yuan}, {Yuan},
  {Yuan}, {Zhai}, {Zhang}, {Zhang}, {Zhang}, {Zhang}, {Zhang}, {Zhang},
  {Zhang}, {Zhang}, {Zhao}, {Zhou}, {Zhou}, {Zhu}, {Zhu}, {Zou}, and
  {Zuo}]{Luo2015}
{Luo}, A.~L., {Zhao}, Y.-H., {Zhao}, G., {Deng}, L.-C., {Liu}, X.-W., {Jing},
  Y.-P., {Wang}, G., {Zhang}, H.-T., {Shi}, J.-R., {Cui}, X.-Q., {Chu}, Y.-Q.,
  {Li}, G.-P., {Bai}, Z.-R., {Wu}, Y., {Cai}, Y., {Cao}, S.-Y., {Cao}, Z.-H.,
  {Carlin}, J.~L., {Chen}, H.-Y., {Chen}, J.-J., {Chen}, K.-X., {Chen}, L.,
  {Chen}, X.-L., {Chen}, X.-Y., {Chen}, Y., {Christlieb}, N., {Chu}, J.-R.,
  {Cui}, C.-Z., {Dong}, Y.-Q., {Du}, B., {Fan}, D.-W., {Feng}, L., {Fu}, J.-N.,
  {Gao}, P., {Gong}, X.-F., {Gu}, B.-Z., {Guo}, Y.-X., {Han}, Z.-W., {He},
  B.-L., {Hou}, J.-L., {Hou}, Y.-H., {Hou}, W., {Hu}, H.-Z., {Hu}, N.-S., {Hu},
  Z.-W., {Huo}, Z.-Y., {Jia}, L., {Jiang}, F.-H., {Jiang}, X., {Jiang}, Z.-B.,
  {Jin}, G., {Kong}, X., {Kong}, X., {Lei}, Y.-J., {Li}, A.-H., {Li}, C.-H.,
  {Li}, G.-W., {Li}, H.-N., {Li}, J., {Li}, Q., {Li}, S., {Li}, S.-S., {Li},
  X.-N., {Li}, Y., {Li}, Y.-B., {Li}, Y.-P., {Liang}, Y., {Lin}, C.-C., {Liu},
  C., {Liu}, G.-R., {Liu}, G.-Q., {Liu}, Z.-G., {Lu}, W.-Z., {Luo}, Y., {Mao},
  Y.-D., {Newberg}, H., {Ni}, J.-J., {Qi}, Z.-X., {Qi}, Y.-J., {Shen}, S.-Y.,
  {Shi}, H.-M., {Song}, J., {Song}, Y.-H., {Su}, D.-Q., {Su}, H.-J., {Tang},
  Z.-H., {Tao}, Q.-S., {Tian}, Y., {Wang}, D., {Wang}, D.-Q., {Wang}, F.-F.,
  {Wang}, G.-M., {Wang}, H., {Wang}, H.-C., {Wang}, J., {Wang}, J.-N., {Wang},
  J.-L., {Wang}, J.-P., {Wang}, J.-X., {Wang}, L., {Wang}, M.-X., {Wang},
  S.-G., {Wang}, S.-Q., {Wang}, X., {Wang}, Y.-N., {Wang}, Y., {Wang}, Y.-F.,
  {Wang}, Y.-F., {Wei}, P., {Wei}, M.-Z., {Wu}, H., {Wu}, K.-F., {Wu}, X.-B.,
  {Wu}, Y.-Z., {Xing}, X.-Z., {Xu}, L.-Z., {Xu}, X.-Q., {Xu}, Y., {Yan}, T.-S.,
  {Yang}, D.-H., {Yang}, H.-F., {Yang}, H.-Q., {Yang}, M., {Yao}, Z.-Q., {Yu},
  Y., {Yuan}, H., {Yuan}, H.-B., {Yuan}, H.-L., {Yuan}, W.-M., {Zhai}, C.,
  {Zhang}, E.-P., {Zhang}, H.-W., {Zhang}, J.-N., {Zhang}, L.-P., {Zhang}, W.,
  {Zhang}, Y., {Zhang}, Y.-X., {Zhang}, Z.-C., {Zhao}, M., {Zhou}, F., {Zhou},
  X., {Zhu}, J., {Zhu}, Y.-T., {Zou}, S.-C., and {Zuo}, F.
\newblock {The first data release (DR1) of the LAMOST regular survey}.
\newblock \emph{Research in Astronomy and Astrophysics}, 15\penalty0
  (8):\penalty0 1095, August 2015.
\newblock \doi{10.1088/1674-4527/15/8/002}.

\bibitem[{Majewski} et~al.(2017){Majewski}, {Schiavon}, {Frinchaboy}, {Allende
  Prieto}, {Barkhouser}, {Bizyaev}, {Blank}, {Brunner}, {Burton}, {Carrera},
  {Chojnowski}, {Cunha}, {Epstein}, {Fitzgerald}, {Garc{\'{\i}}a P{\'e}rez},
  {Hearty}, {Henderson}, {Holtzman}, {Johnson}, {Lam}, {Lawler}, {Maseman},
  {M{\'e}sz{\'a}ros}, {Nelson}, {Nguyen}, {Nidever}, {Pinsonneault},
  {Shetrone}, {Smee}, {Smith}, {Stolberg}, {Skrutskie}, {Walker}, {Wilson},
  {Zasowski}, {Anders}, {Basu}, {Beland}, {Blanton}, {Bovy}, {Brownstein},
  {Carlberg}, {Chaplin}, {Chiappini}, {Eisenstein}, {Elsworth}, {Feuillet},
  {Fleming}, {Galbraith-Frew}, {Garc{\'{\i}}a}, {Garc{\'{\i}}a-Hern{\'a}ndez},
  {Gillespie}, {Girardi}, {Gunn}, {Hasselquist}, {Hayden}, {Hekker}, {Ivans},
  {Kinemuchi}, {Klaene}, {Mahadevan}, {Mathur}, {Mosser}, {Muna}, {Munn},
  {Nichol}, {O'Connell}, {Parejko}, {Robin}, {Rocha-Pinto}, {Schultheis},
  {Serenelli}, {Shane}, {Silva Aguirre}, {Sobeck}, {Thompson}, {Troup},
  {Weinberg}, and {Zamora}]{Majewski2017}
{Majewski}, S.~R., {Schiavon}, R.~P., {Frinchaboy}, P.~M., {Allende Prieto},
  C., {Barkhouser}, R., {Bizyaev}, D., {Blank}, B., {Brunner}, S., {Burton},
  A., {Carrera}, R., {Chojnowski}, S.~D., {Cunha}, K., {Epstein}, C.,
  {Fitzgerald}, G., {Garc{\'{\i}}a P{\'e}rez}, A.~E., {Hearty}, F.~R.,
  {Henderson}, C., {Holtzman}, J.~A., {Johnson}, J.~A., {Lam}, C.~R., {Lawler},
  J.~E., {Maseman}, P., {M{\'e}sz{\'a}ros}, S., {Nelson}, M., {Nguyen}, D.~C.,
  {Nidever}, D.~L., {Pinsonneault}, M., {Shetrone}, M., {Smee}, S., {Smith},
  V.~V., {Stolberg}, T., {Skrutskie}, M.~F., {Walker}, E., {Wilson}, J.~C.,
  {Zasowski}, G., {Anders}, F., {Basu}, S., {Beland}, S., {Blanton}, M.~R.,
  {Bovy}, J., {Brownstein}, J.~R., {Carlberg}, J., {Chaplin}, W., {Chiappini},
  C., {Eisenstein}, D.~J., {Elsworth}, Y., {Feuillet}, D., {Fleming}, S.~W.,
  {Galbraith-Frew}, J., {Garc{\'{\i}}a}, R.~A., {Garc{\'{\i}}a-Hern{\'a}ndez},
  D.~A., {Gillespie}, B.~A., {Girardi}, L., {Gunn}, J.~E., {Hasselquist}, S.,
  {Hayden}, M.~R., {Hekker}, S., {Ivans}, I., {Kinemuchi}, K., {Klaene}, M.,
  {Mahadevan}, S., {Mathur}, S., {Mosser}, B., {Muna}, D., {Munn}, J.~A.,
  {Nichol}, R.~C., {O'Connell}, R.~W., {Parejko}, J.~K., {Robin}, A.~C.,
  {Rocha-Pinto}, H., {Schultheis}, M., {Serenelli}, A.~M., {Shane}, N., {Silva
  Aguirre}, V., {Sobeck}, J.~S., {Thompson}, B., {Troup}, N.~W., {Weinberg},
  D.~H., and {Zamora}, O.
\newblock {The Apache Point Observatory Galactic Evolution Experiment
  (APOGEE)}.
\newblock \emph{\aj}, 154:\penalty0 94, September 2017.
\newblock \doi{10.3847/1538-3881/aa784d}.

\bibitem[Mildenhall et~al.(2020)Mildenhall, Srinivasan, Tancik, Barron,
  Ramamoorthi, and Ng]{Mildenhall2020}
Mildenhall, B., Srinivasan, P.~P., Tancik, M., Barron, J.~T., Ramamoorthi, R.,
  and Ng, R.
\newblock Nerf: Representing scenes as neural radiance fields for view
  synthesis, 2020.

\bibitem[Ness et~al.(2015)Ness, Hogg, Rix, Ho, and Zasowski]{the_cannon}
Ness, M., Hogg, D.~W., Rix, H.-W., Ho, A. Y.~Q., and Zasowski, G.
\newblock The cannon: A data-driven approach to stellar label determination.
\newblock \emph{The Astrophysical Journal}, 808\penalty0 (1):\penalty0 16, jul
  2015.
\newblock \doi{10.1088/0004-637x/808/1/16}.
\newblock URL \url{https://doi.org/10.1088%2F0004-637x%2F808%2F1%2F16}.

\bibitem[{O'Briain} et~al.(2021){O'Briain}, {Ting}, {Fabbro}, {Yi}, {Venn}, and
  {Bialek}]{OBriain2021}
{O'Briain}, T., {Ting}, Y.-S., {Fabbro}, S., {Yi}, K.~M., {Venn}, K., and
  {Bialek}, S.
\newblock {Cycle-StarNet: Bridging the Gap between Theory and Data by
  Leveraging Large Data Sets}.
\newblock \emph{The Astrophysical Journal}, 906\penalty0 (2):\penalty0 130,
  January 2021.
\newblock \doi{10.3847/1538-4357/abca96}.

\bibitem[Parmar et~al.(2018)Parmar, Vaswani, Uszkoreit, Kaiser, Shazeer, Ku,
  and Tran]{Parmar2018}
Parmar, N., Vaswani, A., Uszkoreit, J., Kaiser, L., Shazeer, N., Ku, A., and
  Tran, D.
\newblock Image transformer.
\newblock In \emph{International conference on machine learning}, pp.\
  4055--4064. PMLR, 2018.

\bibitem[{Sajjadi} et~al.(2021){Sajjadi}, {Meyer}, {Pot}, {Bergmann}, {Greff},
  {Radwan}, {Vora}, {Lucic}, {Duckworth}, {Dosovitskiy}, {Uszkoreit},
  {Funkhouser}, and {Tagliasacchi}]{Sajjadi2021}
{Sajjadi}, M. S.~M., {Meyer}, H., {Pot}, E., {Bergmann}, U., {Greff}, K.,
  {Radwan}, N., {Vora}, S., {Lucic}, M., {Duckworth}, D., {Dosovitskiy}, A.,
  {Uszkoreit}, J., {Funkhouser}, T., and {Tagliasacchi}, A.
\newblock {Scene Representation Transformer: Geometry-Free Novel View Synthesis
  Through Set-Latent Scene Representations}.
\newblock \emph{arXiv e-prints}, art. arXiv:2111.13152, November 2021.
\newblock \doi{10.48550/arXiv.2111.13152}.

\bibitem[Shaw et~al.(2018)Shaw, Uszkoreit, and Vaswani]{Shaw2018}
Shaw, P., Uszkoreit, J., and Vaswani, A.
\newblock Self-attention with relative position representations.
\newblock \emph{arXiv preprint arXiv:1803.02155}, 2018.

\bibitem[Sitzmann et~al.(2020)Sitzmann, Martel, Bergman, Lindell, and
  Wetzstein]{Sitzmann2020}
Sitzmann, V., Martel, J. N.~P., Bergman, A.~W., Lindell, D.~B., and Wetzstein,
  G.
\newblock Implicit neural representations with periodic activation functions,
  2020.

\bibitem[Straumit et~al.(2022)Straumit, Tkachenko, Gebruers, Audenaert, Xiang,
  Zari, Aerts, Johnson, Kollmeier, Rix, Beaton, Saders, Teske, Roman-Lopes,
  Ting, and Rom{\'{a} }n-Z{\'{u}}{\~{n}}iga]{Straumit2022}
Straumit, I., Tkachenko, A., Gebruers, S., Audenaert, J., Xiang, M., Zari, E.,
  Aerts, C., Johnson, J.~A., Kollmeier, J.~A., Rix, H.-W., Beaton, R.~L.,
  Saders, J. L.~V., Teske, J., Roman-Lopes, A., Ting, Y.-S., and Rom{\'{a}
  }n-Z{\'{u}}{\~{n}}iga, C.~G.
\newblock Zeta-payne: A fully automated spectrum analysis algorithm for the
  milky way mapper program of the {SDSS}-v survey.
\newblock \emph{The Astronomical Journal}, 163\penalty0 (5):\penalty0 236, apr
  2022.
\newblock \doi{10.3847/1538-3881/ac5f49}.
\newblock URL \url{https://doi.org/10.3847%2F1538-3881%2Fac5f49}.

\bibitem[Ting et~al.(2019)Ting, Conroy, Rix, and Cargile]{the_payne}
Ting, Y.-S., Conroy, C., Rix, H.-W., and Cargile, P.
\newblock The payne: Self-consistent ab initio fitting of stellar spectra.
\newblock \emph{The Astrophysical Journal}, 879\penalty0 (2):\penalty0 69, jul
  2019.
\newblock \doi{10.3847/1538-4357/ab2331}.
\newblock URL \url{https://doi.org/10.3847%2F1538-4357%2Fab2331}.

\bibitem[Vaswani et~al.(2017)Vaswani, Shazeer, Parmar, Uszkoreit, Jones, Gomez,
  Kaiser, and Polosukhin]{Vaswani2017}
Vaswani, A., Shazeer, N., Parmar, N., Uszkoreit, J., Jones, L., Gomez, A.~N.,
  Kaiser, {\L}., and Polosukhin, I.
\newblock Attention is all you need.
\newblock \emph{Advances in neural information processing systems}, 30, 2017.

\bibitem[Wang et~al.(2020)Wang, Li, Khabsa, Fang, and Ma]{Wang2020}
Wang, S., Li, B.~Z., Khabsa, M., Fang, H., and Ma, H.
\newblock Linformer: Self-attention with linear complexity.
\newblock \emph{arXiv preprint arXiv:2006.04768}, 2020.

\bibitem[{Xiang} et~al.(2022){Xiang}, {Rix}, {Ting}, {Kudritzki}, {Conroy},
  {Zari}, {Shi}, {Przybilla}, {Ramirez-Tannus}, {Tkachenko}, {Gebruers}, and
  {Liu}]{Xiang2022}
{Xiang}, M., {Rix}, H.-W., {Ting}, Y.-S., {Kudritzki}, R.-P., {Conroy}, C.,
  {Zari}, E., {Shi}, J.-R., {Przybilla}, N., {Ramirez-Tannus}, M., {Tkachenko},
  A., {Gebruers}, S., and {Liu}, X.-W.
\newblock {Stellar labels for hot stars from low-resolution spectra. I. The
  HotPayne method and results for 330 000 stars from LAMOST DR6}.
\newblock \emph{\aap}, 662:\penalty0 A66, June 2022.
\newblock \doi{10.1051/0004-6361/202141570}.

\bibitem[{Xie} et~al.(2023){Xie}, {Zhang}, {Guo}, {Tan}, {Bian}, {Awadalla},
  {Menezes}, {Qin}, and {Yan}]{2023arXiv230414802X}
{Xie}, S., {Zhang}, H., {Guo}, J., {Tan}, X., {Bian}, J., {Awadalla}, H.~H.,
  {Menezes}, A., {Qin}, T., and {Yan}, R.
\newblock {ResiDual: Transformer with Dual Residual Connections}.
\newblock \emph{arXiv e-prints}, art. arXiv:2304.14802, April 2023.
\newblock \doi{10.48550/arXiv.2304.14802}.

\end{thebibliography}
\bibliographystyle{icml2023}



\end{document}